\documentclass[journal=jacsat,manuscript=article]{achemso}

\usepackage[version=3]{mhchem} % Formula subscripts using \ce{}
\usepackage{glossaries}
\usepackage{nicefrac}
\usepackage{siunitx}
\usepackage{rotating}
\usepackage{multirow}
\usepackage{upgreek}
\usepackage{chemfig}
\usepackage{hyperref}

\author{Christopher Kuenneth}
\author{Arunkumar Chitteth Rajan}
\author{Huan Tran}
\author{Lihua Chen}
\author{Chiho Kim}
\author{Rampi Ramprasad}
\email{rampi.ramprasad@mse.gatech.edu }
\affiliation[Gatech]{School of Materials Science and Engineering, Georgia Institute of Technology, Atlanta, GA 30332, USA}

\title[Multi-Task Learning]
  {Polymer Informatics with Multi-Task Learning}
\keywords{polymer, polymer properties, polymer design, multi-task, machine learning, neural network, Gaussian processing}

\begin{document}

\newacronym{ml}{ML}{machine learning}
\newacronym{mt}{MT}{multi-task}
\newacronym{st}{ST}{single-task}
\newacronym{tg}{$T_\text{g}$}{glass transition temperature}
\newacronym{tm}{$T_\text{m}$}{melting transition temperature}
\newacronym{dft}{DFT}{density functional theory}
\newacronym{fp}{FP}{fingerprint}
\newacronym{rmse}{RMSE}{root mean squared error}
\newacronym{mse}{MSE}{mean squared error}
\newacronym{nn}{NN}{neural network}
\newacronym{pcc}{PCC}{Pearson correlation coefficient}
\newacronym{gpr}{GPR}{Gaussian process regression}
\newacronym{gp}{GP}{Gaussian process}
\newacronym{smiles}{SMILES}{simplified molecular-input line-entry system}
\newacronym{qspr}{QSPR}{quantitative structure-property relationship}
\newacronym{shap}{SHAP}{Shapley additive explanation}
\newacronym{nlp}{NLP}{natural language processing}
 
% \begin{tocentry}

% Some journals require a graphical entry for the Table of Contents.
% This should be laid out ``print ready'' so that the sizing of the

% Inside the \texttt{tocentry} environment, the font used is Helvetica
% 8\,pt, as required by \emph{Journal of the American Chemical

% The surrounding frame is 9\,cm by 3.5\,cm, which is the maximum
% permitted for  \emph{Journal of the American Chemical Society}
% graphical table of content entries. The box will not resize if the
% content is too big: instead it will overflow the edge of the box.

% This box and the associated title will always be printed on a
% separate page at the end of the document.

% \end{tocentry}

\begin{abstract}
Modern data-driven tools are transforming application-specific polymer development cycles. Surrogate models that can be trained to predict the properties of new polymers are becoming commonplace. Nevertheless, these models do not utilize the full breadth of the knowledge available in datasets, which are oftentimes sparse; inherent correlations between different property datasets are disregarded. Here, we demonstrate the potency of multi-task learning approaches that exploit such inherent correlations effectively, particularly when some property dataset sizes are small. Data pertaining to 36 different properties of over  $13, 000$ polymers (corresponding to over $23,000$ data points) are coalesced and supplied to deep-learning multi-task architectures. Compared to conventional single-task learning models (that are trained on individual property datasets independently), the multi-task approach is accurate, efficient, scalable, and amenable to transfer learning as more data on the same or different properties become available. Moreover, these models are interpretable. Chemical rules, that explain how certain features control trends in specific property values, emerge from the present work, paving the way for the rational design of application specific polymers meeting desired property or performance objectives.
\end{abstract}

% \section{Introduction}

Polymers display extraordinary diversity in their chemistry, structure and applications. This is reflected in the ubiquity of polymers in everyday life and technology. The vigor with which polymers are studied using both computational and experimental methods is leading to a constant flux of (mostly uncurated and heterogeneous) data. The field of Polymer Science and Engineering is thus poised for exciting informatics-based inquiry and discovery \cite{Ramprasad2017,Kim2018,Pilania2019}.

In general, materials datasets tend to be small. This presents challenges for the creation of robust and versatile \gls{ml} models for materials property prediction. Nevertheless, the apparent data sparsity in the materials domain is somewhat compensated by the information-richness of each data point or the availability of prior physics-based knowledge of the phenomenon under inquiry. For instance, a given target property A of a material may be correlated with a different property B. If data for A is sparse, but data for B is copious, effective prediction models for A may be developed by exploiting this correlation using algorithms that respect parsimony. Alternatively, imagine that property A may be measured using an accurate (but laborious or expensive) experimental procedure $\alpha$, and a not-so-accurate (but rapid or inexpensive) procedure $\beta$. Again, powerful models for the prediction of property A at the accuracy level of $\alpha$ may be developed by using sparse $\alpha$-type data along with copious $\beta$-type data.

\begin{figure}[hbt]
  \includegraphics[width=\textwidth]{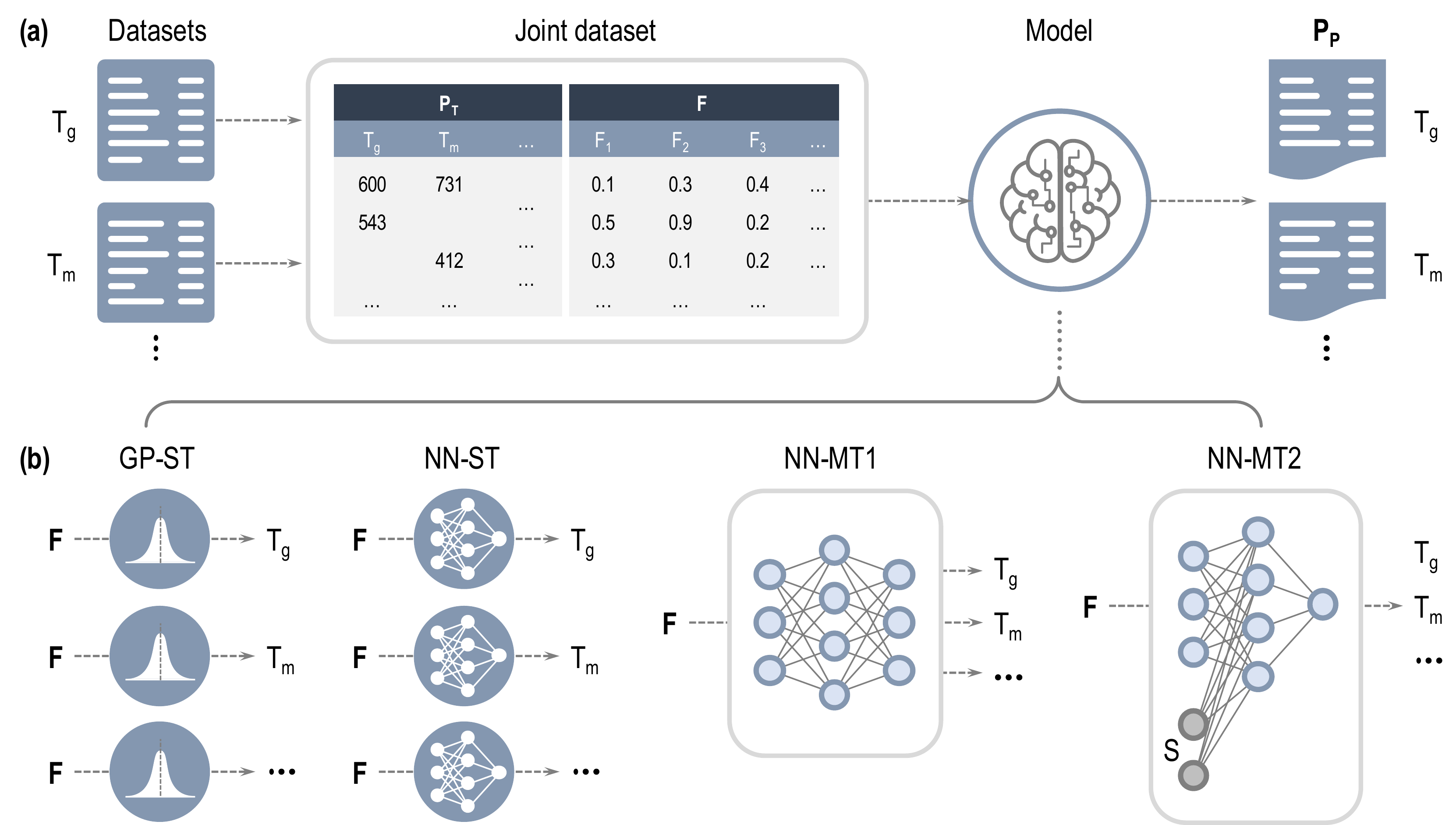}
  \caption{(a) Data pipeline. \emph{From left to right:} Separately collected polymer property datasets are merged into a joint dataset; machine learning models are trained on the joint dataset with fingerprint components ($\mathbf{F}$) as input and predicted properties ($\mathbf{P}_P$) as output. $\mathbf{P}_T$ are the property target values. The loss function is defined as the mean squared error of $\mathbf{P}_P$ and $\mathbf{P}_T$; $T_\text{g}$, $T_\text{m}$, $\mathbf{P}$ and $\mathbf{F}$ stand for glass transition temperature, melting temperature, property and fingerprint component matrix, respectively. (b) Four different machine learning models: \acrfull{st}, \acrfull{mt}, \acrfull{gp}, \acrfull{nn}.}
  \label{fig:scheme}
\end{figure}

With the above in mind, let us suppose that a dataset for a particular materials sub-class involves a variety of target properties, with each property data point potentially obtained from multiple sources or measurements. Not all property values may be available for all material cases. In other words, the dataset may contain a number of ``missing values''. Figure \ref{fig:scheme} (a) shows a schematic of such a dataset for polymer properties. As mentioned above, data from subsets of property and source types may be correlated with each other. Given this scenario, our objective is to utilize a \gls{mt} learning method that can ingest the entire dataset, recognize inherent correlations, and make predictions of all properties by effectively transferring knowledge from one property or source type to another. As a baseline to assess how \gls{mt} learning performs, one may utilize learning methods that learn to predict each property individually, one at a time, implicitly disregarding correlations of the property with other properties; we call these as \gls{st} learning methods. \gls{st} and \gls{mt} learning schemes are illustrated in Figure \ref{fig:scheme} (b).

\gls{mt} learning is an advanced data-driven learning method, which, within materials science, requires coalesced datasets of multiple properties to be effective. While materials scientists have not yet adopted \gls{mt} learning, it has been effectively utilized in drug design for the classification of synthesis-related properties and has demonstrated clear advantages over other learning approaches\cite{Ramsundar2015,Ramsundar2017,Wenzel2019}. A somewhat related approach, which goes under the names of multi-fidelity learning or co-kriging, has been utilized to address some materials science problems \cite{Pilania2017,Batra2019a,Patra2020}; nevertheless, the multi-task learning approach described here surpasses conventional multi-fidelity learning in terms of efficiency, scalability, dataset sizes that can be handled, and the types and number of outputs.

In the present contribution, we focus on polymers and build the first comprehensive \gls{mt} model to-date for the instantaneous prediction of 36 polymer properties. Data for 36 different properties of over $13\,000$ polymers (corresponding to over $23\,000$ data points) were obtained from a variety of sources \cite{Zhu2020,Kim2018,Venkatram2019,Patra2020,Jha2019a,Kim2019,Chen2019,Chen2020}. Table \ref{tbl:properties} shows a synopsis of the data. All polymers are "fingerprinted", i.e., converted to a machine-readable numerical form, using methods described elsewhere\cite{Mannodi-Kanakkithodi2016, Kim2018,Huan2015} (and briefly in the Methods section). These fingerprints (and available property values) are the inputs to our machine learning models. We have developed four types of learning models: two flavors of \gls{mt} models and two flavors of ST models (the latter two models serve as baselines). The two \gls{mt} models utilize \gls{nn} architectures, and are referred to as NN-MT1 and NN-MT2 models. Once trained on the coalesced datasets corresponding to 36 polymer properties, the NN-MT1 model takes in polymer fingerprints for a new polymer and outputs all 36 properties via its last multi-head output layer. The NN-MT2 model, on the other hand, uses an architecture that receives the concatenation of the new polymer fingerprint and a selector vector as input. The selector vector indicates the property/source, and instructs the NN to output just that selected property/source. The baseline ST models utilize either Gaussian processes (GP-ST) or a conventional NN architecture (NN-ST). The GP-ST and NN-ST models are trained independently on individual polymer datasets; there are thus 36 prediction models, one for each property, of each ST flavor. All four ML approaches developed here are shown in Figure \ref{fig:scheme} (b); details on the architecture of the models and training process are provided in the Methods section. 
\section{Results}

\subsection{Correlations in Data}

\begin{figure}[hbpt]
  \includegraphics[width=\textwidth]{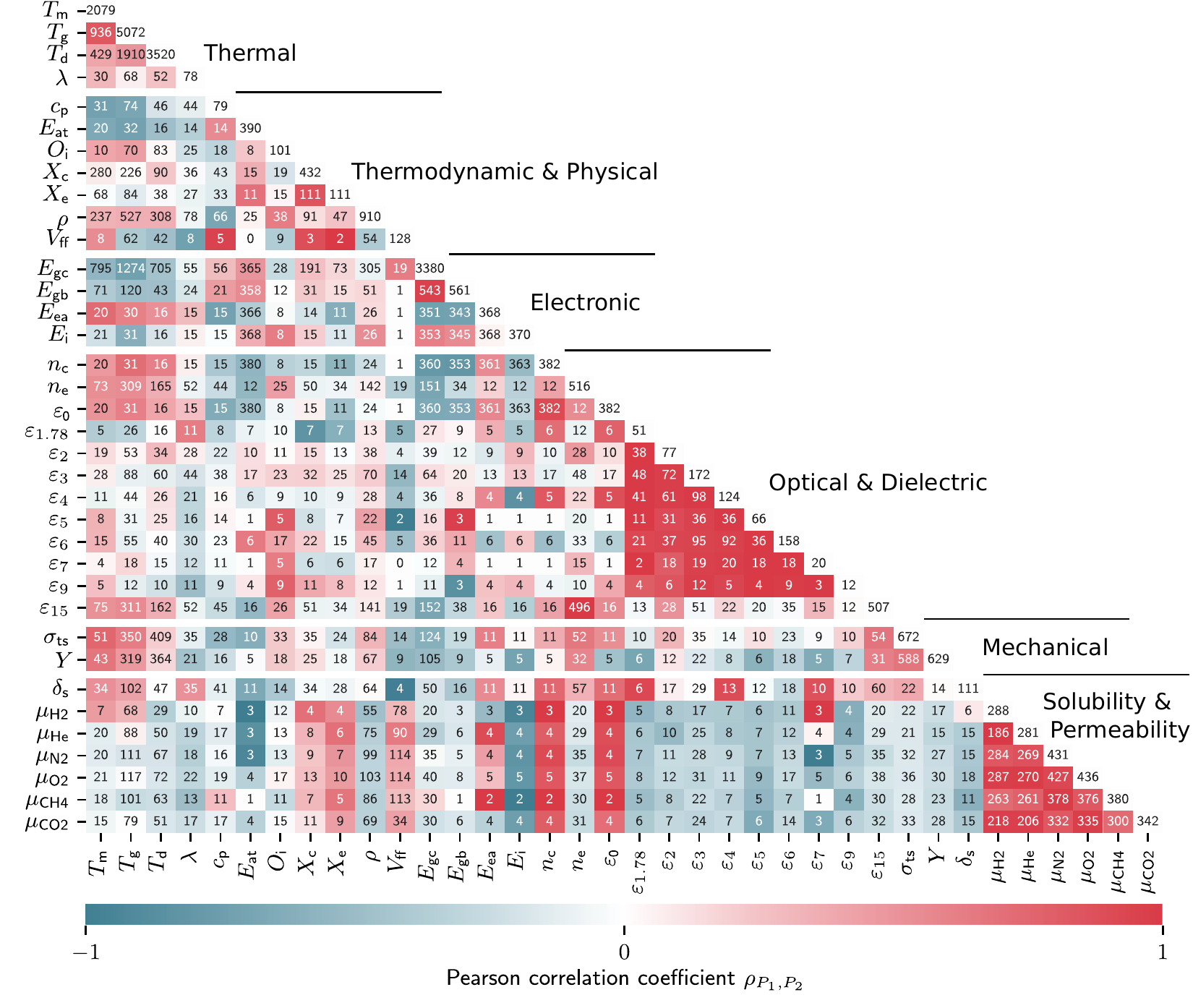}
  \caption{Polymer property heatmap of the Pearson correlation coefficients (PCCs). Red patches indicate positively correlated, blue patches negatively correlated, and white patches uncorrelated PCCs. Off-diagonal numbers denote the congruences between two polymer properties (polymers provided in both datasets), whereas diagonal numbers indicate the total count of a particular property. The PCCs for less than two congruences were set to 0. Property symbols are defined in Table \ref{tbl:properties}.}
  \label{fig:pearson_heatmap}
\end{figure}

Unlike \gls{st} models, \gls{mt} models learn from inherent correlations in datasets. Our polymer dataset shows such interesting (some expected, but some new) correlations between pairs of properties as illustrated in Figure \ref{fig:pearson_heatmap} using \glspl{pcc}. For example, the dielectric constants at different frequencies (ranging from $\varepsilon_{1.78}$ to $\varepsilon_{9}$, where the subscript indicates the frequency on $\log_{10}$ scale in Hz) are highly positively correlated with each other. Understandably, the dielectric constant at optical frequency, $\varepsilon_{15}$, which is controlled purely by electronic polarization, is weakly correlated with the dielectric constants at low frequencies, which are related to ionic, orientational and electronic factors. The permeabilities of gases, $\mu_\text{g}$ (where g represents one of 6 gas molecules), are highly positively correlated  with each other. By contrast, gas permeabilities and dielectric constants are negatively correlated with each other, indicating that polymers with high $\varepsilon_f$ tend to display low $\mu_g$, and \emph{vice versa}. Of note, high positive correlations can be seen between the glass transition ($T_\text{g}$) and melting temperatures ($T_\text{m}$), and large negative correlation between the electronic band gap ($E_\text{gb}$, for bulk polymers, and $E_\text{gc}$, for chains) and $T_\text{g}$. The important observation that should be made by the inspection of Figure \ref{fig:pearson_heatmap} is that there are several examples of weak to strong positive and negative correlations between properties that can potentially be exploited in MT learning schemes.

\begin{table}[hbtp]
  \caption{Synopsis of 36 polymer properties. The total number of single data points is $23\,616$ and the total number of merged data points in the joint database is $13\, 766$.}
  \label{tbl:properties}
\begin{tabular}{lllllll}
\hline
Property                     & Symbol        & Unit               & Source\textsuperscript{\emph{a}} & Points & Data Range          & Ref.                               \\ \hline
\multicolumn{7}{c}{\textbf{Thermal}} \\
Melting & \multirow{2}{*}{$T_\text{m}$}  & \multirow{2}{*}{\si{K}}             & \multirow{2}{*}{Exp.}   & \multirow{2}{*}{2079}        & \multirow{2}{*}{$[226, 860]$}         & \multirow{2}{*}{ } \\
temperature & & & & & & \\

Glass transition & \multirow{2}{*}{$T_\text{g}$}  & \multirow{2}{*}{\si{K}}             & \multirow{2}{*}{Exp.}   & \multirow{2}{*}{5072}        & \multirow{2}{*}{$[80, 873]$}         & \citenum{Jha2019a,Kim2019} \\
temperature & & & & & & \citenum{Kim2018} \\

Decomposition & \multirow{2}{*}{$T_\text{d}$}  & \multirow{2}{*}{\si{K}}             & \multirow{2}{*}{Exp.}   & \multirow{2}{*}{3520}        & \multirow{2}{*}{$[219, 1167]$}         & \multirow{2}{*}{ } \\
temperature & & & & & & \\

Thermal conductivity                & $\lambda$  & \si{W\,mK^{-1}} & Exp.   & 78          & $[0.1, 0.49]$       &    \\

\multicolumn{7}{c}{\textbf{Thermodynamic \& Physical}} \\

Heat capacity                & $c_\text{p}$  & \si{J\,gK^{-1}} & Exp.   & 79          & $[0.8, 2.1]$       &                                    \\
Atomization energy           & $E_\text{at}$         & \si{eV\,atom^{-1}}       & DFT    & 390      & $[-6.8, 5.2]$     & \citenum{Kim2018}  \\

Limiting oxygen index & $O_\text{i}$  & \si{\%} & Exp.   & 101  &  $[13.2, 70]$       &    \\

Crystallization tendency (DFT)  & $X_\text{c}$  & \si{\%}   & DFT    & 432         & $[0.1, 98.8]$ &                                    \\
Crystallization tendency (exp.)   & $X_\text{e}$  & \si{\%}   & Exp.    & 111         & $[1, 98.5]$ &                                    \\

Density                      & $\rho $       & \si{g\,cm^{-3}} & Exp.   & 910  & $[0.84, 2.18]$      & \citenum{Kim2018}                  \\
Fractional free volume & $V_\text{ff}$  & 1 & Exp.   & 128  &  $[0.1, 0.47]$       &    \\

\multicolumn{7}{c}{\textbf{Electronic}} \\

Bandgap (chain)             & $E_\text{gc}$  & \si{eV}            & DFT    & 3380         & $[0.02, 9.86]$ &                                    \\
Bandgap (bulk)              & $E_\text{gb}$  & \si{eV}            & DFT    & 561        & $[0.4, 10.1]$  & \citenum{Patra2020}                \\

Electron affinity          & $E_\text{ea}$      & \si{eV}            & DFT    & 368         & $[-0.39, 5.17]$  &                                    \\
Ionization energy            & $E_\text{i}$         & \si{eV}            & DFT    & 370         & $[3.56, 9.84]$     &                                    \\

\multicolumn{7}{c}{\textbf{Optical \& Dielectric}} \\
Refractive index (DFT) & $n_\text{c}$  & \si{1}             & DFT & 382    & $[1.48, 2.95]$  & \citenum{Kim2018}                  \\
Refractive index  (exp.)& $n_\text{e}$  & \si{1}             & Exp. & 516     & $[1.29, 2]$  & \citenum{Kim2018}                  \\

Dielectric constant          & $\varepsilon_0$ & \si{1}             & DFT   & 382       & $[2.6, 9.1]$      & \citenum{Kim2018}                  \\

Frequency dependent & \multirow{2}{*}{$\varepsilon_f$ \textsuperscript{\emph{b}} }  & \multirow{2}{*}{1} & \multirow{2}{*}{Exp.}   &    \multirow{2}{*}{1187}      &    \multirow{2}{*}{[1.95, 10.4]}    &     \citenum{Chen2020}  \\
dielectric constant  & & & & & & \\

\multicolumn{7}{c}{\textbf{Mechanical}} \\

Tensile strength & $\sigma_\text{ts}$  & \si{MPa} & Exp.   & 672 & $[2.86, 289]$       &    \\
Young's modulus & $Y$  & \si{MPa} & Exp.   & 629  & $[0.02, 9.8]$       &    \\

\multicolumn{7}{c}{\textbf{Solubility \& Permeability}} \\

Hildebrand solubility & \multirow{ 2}{*}{$\delta_\text{s}$}      & \multirow{ 2}{*}{\si{\sqrt{MPa}}}   & \multirow{ 2}{*}{Exp.}   & \multirow{2}{*}{112} & \multirow{ 2}{*}{$[12.3, 29.2]$}      & \multirow{ 2}{*}{\citenum{Venkatram2019, Kim2018}}   \\
parameter & & & & & & \\

Gas permeability & $\mu_g$ \textsuperscript{\emph{c}} & Barrer & Exp. & 2168 & [0, 4.7]\textsuperscript{\emph{d}} & \citenum{Zhu2020}  \\

\hline
\end{tabular}
\begin{flushleft}
\textsuperscript{\emph{a}} Experiments (Exp.); density functional theory (DFT)

\textsuperscript{\emph{b}} $f \in \left\{1.78, 2, 3, 4, 5, 6, 7, 9, 15 \right\}$ is the $\log_{10}$ (frequency in Hz); e.g., $\varepsilon_3$ is the dielectric constant at a frequency of \SI{1}{kHz}

\textsuperscript{\emph{c}} $g \in \left\{\text{\ce{He}, \ce{H2}, \ce{CO2}, \ce{O2}, \ce{N2}, \ce{CH4}}\right\}$

\textsuperscript{\emph{d}} The data range is transformed by $f: \mu_\text{g} \mapsto \log_{10} \left( \mu_\text{g} + 1 \right)$
\end{flushleft}
\end{table}

\subsection{Single and Multi-Task Models}

\begin{figure}[hbt]
  \includegraphics[width=1\textwidth]{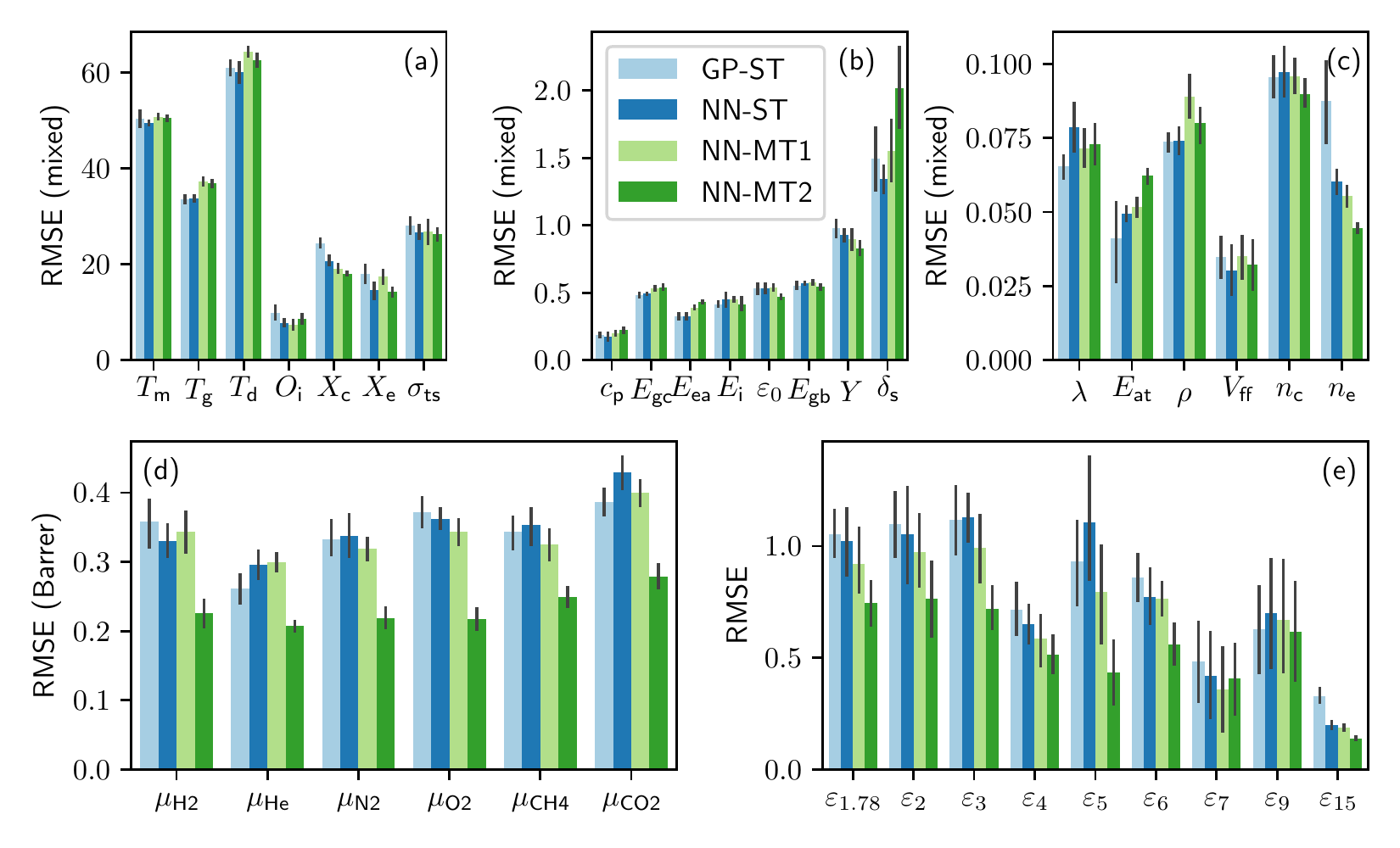}
  \caption{Five-fold cross validation root mean squared errors (RMSEs) of four machine learning models for 36 polymer properties. The properties are arranged in sub-figures according to their magnitudes. The colored bars indicate the average of five-fold cross validation RMSEs, and the error bars are the \SI{68}{\%} confidence intervals of the RMSE averages. Units, can be found in Table \ref{tbl:properties}.}
  \label{fig:comparison}
\end{figure}

To investigate whether these correlations improve the prediction performance when used in MT models, we train four different ML models. The first two models use the ST architecture, which predicts single polymer properties, and the next two models use the MT architecture, which predicts all properties (see Figure \ref{fig:scheme} (b)). Figure \ref{fig:comparison} compiles the training results of all four models. The average of the five-fold cross-validation \glspl{rmse} of the unseen validation dataset are shown, with the error bars indicating \SI{68}{\%} confidence intervals of the \gls{rmse} averages. A more condensed overview of the training results, using categories defined in Figure \ref{fig:pearson_heatmap}, is shown in Table \ref{tbl:overall_comp}.

% ST vs ST
Using the training results in Figure \ref{fig:comparison} and Table \ref{tbl:overall_comp}, we first evaluate the performance of both ST models (GP-ST and NN-ST). In general, we find the NN-based \gls{st} models to perform better than their \gls{gp} counterparts. This is an interesting result as both models only differ in their underlying learning algorithm but otherwise follow the same \gls{st} doctrine that learns polymer properties independently. Nevertheless, it is also known that NN models can approximate more general function classes better than \gls{gp} models\cite{Csji2001ApproximationWA}, which ultimately leads to the better overall performance of the NN models. 

% ST vs MT
Next, we compare the ST and MT models using the average of the normalized RMSE values in Table \ref{tbl:overall_comp}. The RMSE values were normalized so that the maximal value is 1. Overall, Table \ref{tbl:overall_comp} shows that the MT models perform generally better than the ST models. The NN-MT2 model provides the best accuracy over all the 36 properties and is followed by NN-MT1, which is superior to NN-ST (with GP-ST finishing up last). Comparing the first three categories (thermal, thermodynamic \& physical, and electronic) of Table \ref{tbl:overall_comp}, the ST models perform slightly better than the MT models. Similar observations can be made in Figures \ref{fig:comparison} (a), (b), and (c), which comprise the properties of these first three categories. The reason for the good performance of the ST models on these three categories is the copious amount of data that is available. This allows the optimizer to fully focus on a large single-property space during the optimization. MT models on the other hand tend to compromise their performance on these cases to also be able to provide high predictive performance for the cases where the dataset is sparse (where ST models suffer). Moreover, given that the property values are scaled to similar data ranges in the pre-processing step (see Method section), the MT models are effectively trained on data ranges with different sparsities, which present numerical challenges for the optimizer. The last three categories in Table \ref{tbl:overall_comp} and Figures \ref{fig:comparison} (d) and (e) paint a different picture compared to the first three categories; the MT models clearly outperform the ST models. The last three categories comprise the highly correlated properties $\mu_g$ and $\varepsilon_f$ (c.f. Figure \ref{fig:pearson_heatmap}) and also those with small datasets, which the MT models use to improve their prediction performance. We note that although Table \ref{tbl:properties} indicates that there are 1187 points for the frequency-dependent dielectric constant, this dataset is spread across 10 frequencies.

Among the MT models, the concatenation-based NN-MT2 model displays a significantly lower averaged RMSE of 0.8. The degraded performance of the multi-head NN-MT1 model in comparison to NN-MT2 may be ascribed to the sparse population of our dataset due to missing properties for many polymers. When the optimizer computes the gradients to back-propagate over the network, it has to exclude these missing properties, effectively leaving related network parts unchanged. The architecture of the NN-MT2 eliminates this problem by using a one-hot representation of the dataset. As this representation has no missing values, the optimizer can always back-propagate over the entire network. 

\begin{table}[hbtp]
  \caption{The average of normalized RMSE values. RMSE averages were normalized for each property so that the maximal value is 1.}
  \label{tbl:overall_comp}
\begin{tabular}{lc|cccccc}
\hline
\multirow{3}{*}{Model} & \multirow{3}{*}{All} & \multicolumn{6}{c}{Categories}  \\
    &                                        & Thermal & Thermod. & Electronic & Optical & Mech. & Solubility  \\
      &         &         & \& Physical   &            & \& Dielectric &      &  \& Permeability \\
\hline
GP-ST  &  0.94 &              \textbf{0.92} &       0.90 &        \textbf{0.88} &     0.98 &        1.00 &  0.93 \\
NN-ST  &  0.90 &              0.95 &       \textbf{0.82} &        0.91 &     0.91 &        0.95 &  0.94 \\
NN-MT1 &  0.89 &              0.98 &         0.89        &        0.97 &     0.83 &        0.94 &  0.92 \\
NN-MT2 &  \textbf{0.80} &     0.97  &               0.89 &        0.96 &     \textbf{0.69} &        \textbf{0.89} &  \textbf{0.70} \\
\hline
\end{tabular}
\end{table}

% All
By holistically evaluating the performance of all properties and models in Figure \ref{fig:comparison} and Table \ref{tbl:overall_comp}, it can be stated that \glspl{nn} should be preferred over \glspl{gp}. \glspl{nn} predict not only with higher accuracy than \glspl{gp} but they also scale efficiently in terms of growing dataset, training and prediction time. \gls{mt} models comprise similar accuracy than ST models and should particularly be utilized whenever the considered data exhibit high correlations. Finally, the NN-MT2 model displays the overall best performance among all properties for our dataset.

\subsection{Deriving Chemical Guidelines}

\begin{figure}[hbt]
  \includegraphics{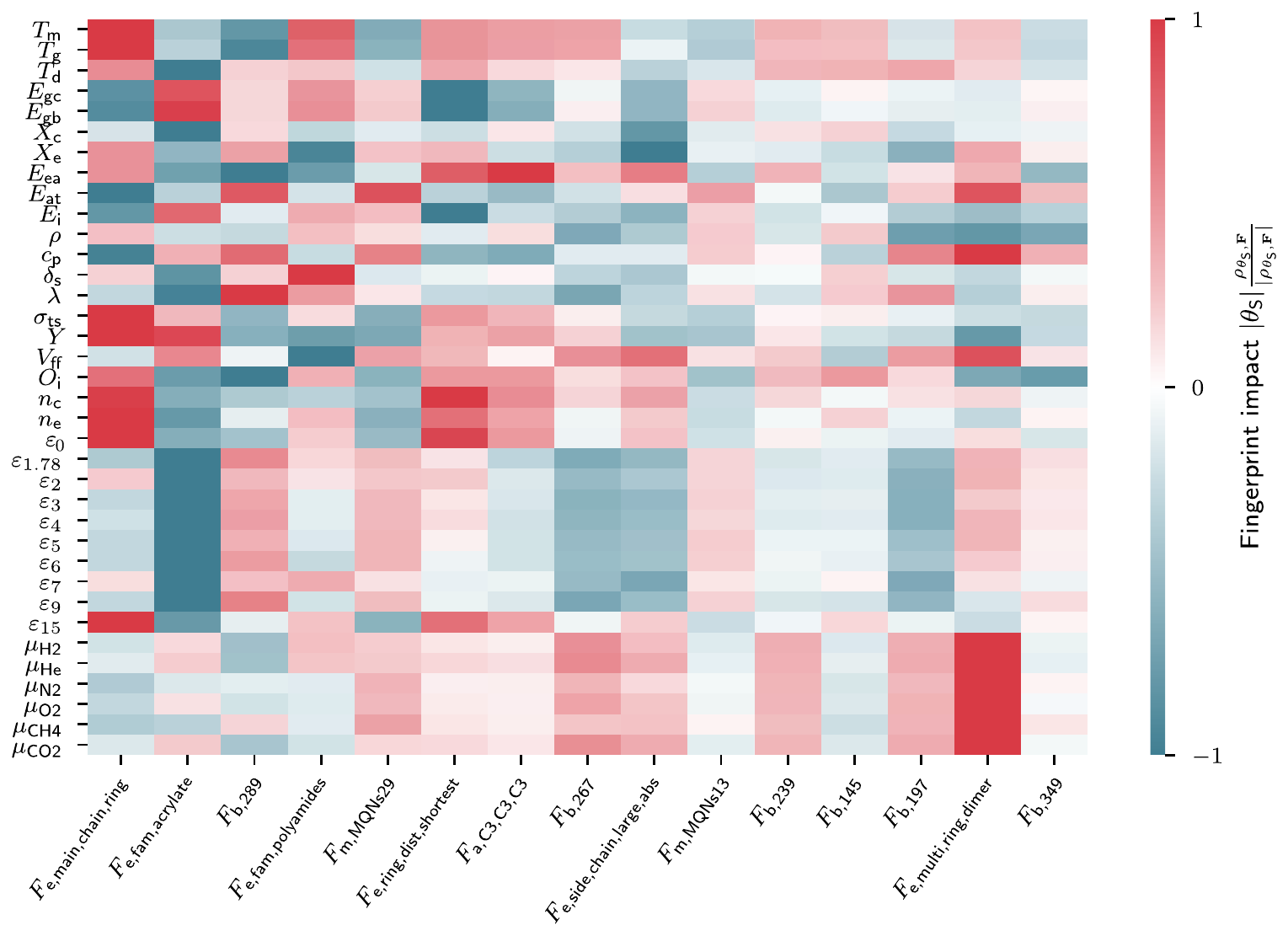}
  \caption{Fingerprint impact values of the 15 most important fingerprint components. Positive fingerprint impact values (red) indicate that a positive value of a fingerprint will potentially increase a property value, and \emph{vice versa}. Small impact values, however, suggest little or no change of the property value.}
  \label{fig:shap}
\end{figure}

% The fingerprint impact values are computed as the product of the \acrfull{shap}  values $|\theta_\text{S}|$ and sign of the \acrfull{pcc} $\rho_{\theta_\text{S}, \mathbf{F}}$ of the SHAP values and the fingerprint values $|\theta_\text{S}| \cdot \nicefrac{\rho_{\theta_\text{S}, \mathbf{F}}} {|\rho_{\theta_\text{S}, \mathbf{F}}|}$. Thus, positive fingerprint impact values ($\propto$) indicate that a large fingerprint value increase the property and \emph{vice versa}, whereas the impact values' magnitude denotes the the responsibility (strength) of the impact.

While our ML models learn the mapping between fingerprints of polymers and properties, they do not provide insights on how single fingerprint components relate to properties, nor how modifications of fingerprint components affect properties. To address such problems, we calculate \Gls{shap} values\cite{Lundberg2017,Shrikumar2017} that measure the impact of structural polymer features, which are indicated through our fingerprint components, to polymer properties. As these \gls{shap} values do only quantify the magnitude of the fingerprint-structure relation and not the direction, we compute special fingerprint impact values as the product of \gls{shap} and \gls{pcc} values. Using these fingerprint impact values, we derive chemical guidelines that can be compared with well-known empirical guidelines from the literature to validate our model irrespective of the used training dataset. 

The most influential fingerprint component in Figure \ref{fig:shap} is $F_\text{e,main,chain,ring}$, which is defined as the ratio of the number of non-hydrogen atoms in rings (cycles of atoms) to the total number of atoms, and is large for polymers containing many rings. $F_\text{e,main,chain,ring}$ has a strong positive impact on $T_\text{m}$, $T_\text{g}$, $T_\text{d}$, $\sigma_\text{ts}$, $Y$, $O_\text{i}$, $n_e$, $n_c$, $\varepsilon_0$ and $\varepsilon_{15}$, and strong negative impact on $E_\text{gc}$, $E_\text{gb}$, $E_\text{at}$, $E_\text{i}$ and $c_\text{p}$. This means, the presence of atomic rings increase the former-mentioned properties but decrease the latter. As such, using the fingerprint impacts, we can provide chemical guidelines helpful to design future polymers. The derived chemical guidelines as impacts of the fingerprint component $F_{e,main,chain,ring}$ may be mapped to empirical guidelines that scientists have learned over the years. For instance, it is known that the presence of atomic rings stiffens polymers, which explains the increase of the mechanical properties, $\sigma_\text{ts}$ and $Y$. Moreover, the atomic rings restrict chain motion, which is the reason for increased $T_\text{m}$, $T_\text{g}$ and $T_\text{d}$ values in ring-rich polymers. The conjugated double bonds in atomic rings introduce agitated $\pi$-electrons, which increase $n_e$, $n_c$ and $\varepsilon_f$, especially at high frequencies ($\varepsilon_{15}$) where electronic displacements contribute significantly to optical properties. Also, the agitated $\pi$-electrons of atomic rings can participate in electrical conduction, which is why rings increase the conductivity of polymers. In contrast, properties such as $E_\text{gc}$, $E_\text{gb}$, $E_\text{at}$, and $E_\text{i}$ \cite{van2009properties}, which correlate with insulating behavior or stability, are decreased as $F_\text{e,main,chain,ring}$ has negative impact. 

The second-most impactful fingerprint component is $F_\text{e,fam,acrylate}$, which is defined to be one if the acrylate group is present in the polymer and zero otherwise. Polyacrylates are known to have $T_\text{g}$ values below room temperature. Consistent with this expectation, the presence of $F_\text{e,fam,acrylate}$ negatively impacts $T_\text{m}$, $T_\text{d}$ and $T_\text{g}$. Another interesting finding is that $F_\text{e,fam,polyamides}$, the fourth-most impactful fingerprint component, positively impacts $\delta_\text{s}$ because the amide bonds in polyamides strengthen inter-molecular forces that make polymers resist dissolution. One can likewise derive useful insights from the other features identified in Figure \ref{fig:shap}.

\section{Discussion}

In this work, we demonstrate how MT learning improves the property prediction of ML models in materials sciences by using inherent property correlations of coalesced datasets. Our polymer dataset includes 36 properties from over $23\,000$ data points of more than $13\,000$ polymers. The dataset is learned using four different ML models: the first two models are based on \glspl{gp} and \glspl{nn} and use the ST architecture. Model three and four are solely based on \glspl{nn} and use two different types of MT architectures. Our analysis shows that the fourth model (NN-MT2) outperforms the other three models overall. Upon closer inspection of performance within individual property sub-classes, it is evident that MT models outperform ST models especially when correlations between properties within the subclass is high and/or when the dataset sizes within those sub-classes is small. In closing, we conclude that MT learning successfully improved the property prediction by utilizing the inherent correlations in our coalesced polymer dataset. Furthermore, we compute fingerprint impact values, which are based on \gls{shap} and \gls{pcc} values, that allow us to derive chemical guidelines for polymer design from the trained MT model and adds an additional validation (and value-added) step pertaining to knowledge extraction.

Besides better performance, our MT learning approach makes fast predictions of all properties in a short time and eliminates the laborious training of many single ML models for each property. In addition, the NNs enable scalability and fast retraining of the \gls{mt} models when new properties or data become available. \gls{mt} models can also be developed further to include uncertainty quantifications, which is often helpful for end-users. Also, it is important to note that our MT learning approach and fingerprint impact analysis are not limited to polymeric materials; in fact, they can easily be modified to handle any material. Given all these factors and the good performance, we believe that MT models should be the preferred method for property predictive ML in materials informatics. All ML models developed in this work will be made available on the Polymer Genome platform at \url{www.polymergenome.org}.

\section{Methods}
\subsection{Data and Preparation}

The polymer database used in this work comprises 36 individual polymer properties, which are meticulously collected and curated from two main sources: (i) in-house high-accuracy and high-throughput \gls{dft} based computations\citenum{Huan2015,Huan2016,Patra2020,Sharma2014}, (ii) experimental measurements reported in the literature (as referenced in Table \ref{tbl:properties}), printed handbooks\cite{J.Brandrup1999,Barton1991,Bicerano2002} and online databases.\cite{polymerdatabase,polyinfo}. Both sources come with distinct uncertainties and should not be mixed together under the same property; while DFT contains systematic uncertainties introduced through the approximations of the density functional or chosen convergence parameters, experimental uncertainties arise from sample and measurement conditions. However, along the lines of multi-fidelity learning approaches, the concurrent use of properties of different sources in separate columns of the dataset may help to lower the total generalization error. An overview of the used 36 polymer properties, their symbols, units, sources, and data ranges can be found in Table \ref{tbl:properties}. It should be noted that some of the individual property datasets have already been used in other publications (see references in Table \ref{tbl:properties}). However, this work marks the first time that these single property datasets have been fused for the holistic training of the MT models.

\gls{mt} architectures take in the fingerprint of a polymer and use the same \gls{nn} to predict on all properties. The property prediction happens either simultaneously, as in our multi-head MT model (NN-MT1), or iteratively, as in our concatenation-based MT model (NN-MT2). \gls{mt} models thus need a coalesced dataset that lists all properties for one polymer per row, see Figure \ref{fig:scheme} (a). To construct such a coalesced dataset, we merge the 36 single-property datasets using our polymer fingerprints (see next Section). Moreover, to ease the work of the optimizer and accelerate the training, we scale all 36 property values to a comparable data range using Scikit-learn's Robust Scaler\cite{Varoquaux2015}. Additionally, the gas permeabilities ($\mu_\text{g}$) are logartihmically pre-processed by $f: \mu_\text{g} \mapsto \log_{10} \left( \mu_\text{g} + 1 \right)$ to narrow down their large data range. Ultimately, for computing error measurements and production, the original metrics are restored by inversely transforming the predictions. Apart from scaling the polymer properties, fingerprint components are normalized to the range of $[0, 1]$. 

\subsection{Fingerprinting}

Fingerprinting converts geometric and chemical information of polymers to machine-readable numerical representations. Polymer chemical structures are represented using SMILES\cite{Weininger1988} strings that follow the SMILES syntax but use two stars to indicate the two endpoints of the repetitive unit of the polymers. 

Our polymer fingerprints capture key features of polymers at three hierarchical length scales\cite{Mannodi-Kanakkithodi2016}. At the atomic-scale, our fingerprints track the occurrence of a fixed set of atomic fragments (or motifs)\cite{Huan2015,Mannodi-Kanakkithodi2017a}. For example, the fragment ``\verb=O1-C3-C4='' is made up of three contiguous atoms, namely, a one-fold coordinated oxygen, a 3-fold coordinated carbon, and a 4-fold coordinated carbon, in this order. A vector of such triplets form the fingerprint components at the lowest hierarchy. The next level uses the \gls{qspr} fingerprints\cite{Le2012, Kim2018} to capture features on larger length-scales. \gls{qspr} fingerprints are often used in chemical and biological sciences, and implemented in the cheminformatics toolkit RDKit\cite{rdkit}. Examples of such fingerprints are the van der Waals surface area\cite{Iler1995}, the topological polar surface area (TPSA)\cite{Ertl2000,Prasanna2008}, the fraction of atoms that are part of rings (i.e., the number of atoms associated with rings divided by the total number of atoms in the formula unit), and the fraction of rotatable bonds. The highest length-scale fingerprint components in our polymer fingerprints deal with “morphological descriptors”. They include features such as the shortest topological distance between rings, fraction of atoms that are part of side-chains, and the length of the largest side-chain. Eventually, the used polymer fingerprint vector ($\mathbf{F}$) of a polymer in this study has 953 components of which 371 are from the first, 522 from the second and 60 from the third level.

\subsection{Machine Learning Models}

To allow for comparison our four \gls{ml} models, we consistently chose the loss function being the \gls{mse} of predicted and true values for five different training datasets, generated by five-fold cross-validation. The five-fold cross validation means along with the \si{68\%} confidence intervals are reported in Figure \ref{fig:comparison}.

\subsubsection{Single-Task Learning with Gaussian Process Regression (GP-ST)}

Scikit-learn's\cite{Varoquaux2015} implementation of \gls{gp} regression was used as the baseline model, denoted by GP-ST. The kernel function was chosen as the parameterized radial basis functions plus a white kernel contribution to capture noise. \gls{gp} predicts probability distributions from which prediction values are derived as the means of the distributions, and confidence intervals of the distributions define the uncertainties. \gls{gp}'s limiting factor is the inversion of the kernel matrix, which grows squared ($\mathbf{F}^2$) with the number of used features ($\mathbf{F}$), rendering \gls{gp} unsuitable for big-data learning problems. NNs eliminate this problem.

\subsubsection{Learning with Neural Networks (NN-ST, NN-MT1, NN-MT2)}

All three \gls{nn} models were implemented using the Python API of Tensorflow\cite{tensorflow}. We used the Adam optimizer with a learning rate of $10^{-3}$ to minimize the \gls{mse} of the prediction and target polymer property. Early stopping combined with a learning rate scheduler was deployed. All hyper-parameters such as the initial learning rate, number of layers and neurons were optimized with respect to the generalization error using the Hyperband method\cite{Li2018} of the Python package Keras-Tuner.

The NN-ST model takes in the fingerprint vector and outputs one polymer property. Just as the GP-ST model, we train an ensemble of 36 independent NN-ST models to predict all 36 properties. The NN-MT1 model has a multi-head MT architecture that takes in the fingerprint vector ($\mathbf{F}$) and outputs 36 properties ($\mathbf{P}$) at the same time. On the other hand, the NN-MT2 model uses a concatenation-based MT architecture that takes in the fingerprint vector and a selector vector $\mathbf{S}$, outputting only the selected polymer property. The selector vector has 36 components where one component is \texttt{1} and the rest \texttt{0}. Each of the three NN models has two dense layers, followed by a parameterized ReLU activation function and a dropout layer with rate 0.5. The Hyperband method optimized the two dense layers to 480 and 224 neurons for the NN-ST model, 480 and 416 neurons for the NN-MT1 model, and 224 and 160 neurons for NN-MT2 model. An additional dense layer was added with 1 neuron for the NN-ST and NN-MT2 model and 36 for the NN-MT2 model to resize the output layer. 

\subsection{SHAP}

The Shapley's cooperative game theory-based SHAP (SHapley Additive exPlanations)\cite{Lundberg2017,Shrikumar2017} analysis is a unified framework for interpreting predictions of ML models by assigning impact values to input features. To establish the interpretability of fingerprint components and polymer properties in our work, we intially compute \gls{shap} values for the prediction on the validation dataset using the best NN-MT1 model. Since these raw \gls{shap} values ($\theta_\text{S}$) indicate a fingerprint's ability to amend certain polymer properties, mean sums of absolute \gls{shap} values $|\theta_\text{S}|$ may be used to measure the total fingerprint component impact on each property. However, $|\theta_\text{S}|$ does not measure the proportionality of fingerprint and property, that is to say, the positive or negative change of the property owing to the fingerprint. This is why we compute the \glspl{pcc} of \gls{shap} and fingerprint components, $\rho_{\theta_\text{S}, \mathbf{F}}$, and multiply these PCC with the mean sum of the absolute \gls{shap} values, finally leading to our definition of the fingerprint impact values as $|\theta_\text{S}| \cdot \nicefrac{\rho_{\theta_\text{S}, \mathbf{F}}} {|\rho_{\theta_\text{S}, \mathbf{F}}|}$. \gls{shap} values were computed using the GradientExplainer class of the \gls{shap} python package (\url{https://github.com/slundberg/shap}).

\begin{acknowledgement}

C. Ku. thanks the Alexander von Humboldt Foundation for financial support. This work is financially supported by the Office of Naval Research through a Multi-University Research Initiative (MURI) grant (N00014-17-1-2656)  and a regular grant (N00014-20-2175).

\end{acknowledgement}

\section{Author Contributions}
C.Ku. designed, trained and evaluated the machine learning models and wrote this paper. L.C., H.T., A.C.R. and C.Ki. collected and curated the polymer property database. The work was conceived and guided by R.R. All authors discussed results and commented on the manuscript.

% \begin{suppinfo}

% \end{suppinfo}

\bibliography{references}

\end{document}